\documentstyle[twoside]{article}

\catcode`\@=11
\long\def\@makefntext#1{
\protect\noindent \hbox to 3.2pt {\hskip-.9pt  

$^{{\eightrm\@thefnmark}}$\hfil}#1\hfill}               

\def\thefootnote{\fnsymbol{footnote}}
\def\@makefnmark{\hbox to 0pt{$^{\@thefnmark}$\hss}}    

\def\ps@myheadings{\let\@mkboth\@gobbletwo
\def\@oddhead{\hbox{}
\rightmark\hfil\eightrm\thepage}   

\def\@oddfoot{}\def\@evenhead{\eightrm\thepage\hfil
\leftmark\hbox{}}\def\@evenfoot{}
\def\sectionmark##1{}\def\subsectionmark##1{}}

\oddsidemargin=\evensidemargin
\renewcommand{\thefootnote}{\fnsymbol{footnote}}

\newcounter{sectionc}\newcounter{subsectionc}\newcounter{subsubsectionc}
\renewcommand{\section}[1] {\vspace{12pt}\addtocounter{sectionc}{1} 

\setcounter{subsectionc}{0}\setcounter{subsubsectionc}{0}\noindent 

        {\tenbf\thesectionc. #1}\par\vspace{5pt}}
\renewcommand{\subsection}[1] {\vspace{12pt}\addtocounter{subsectionc}{1} 

        \setcounter{subsubsectionc}{0}\noindent 

        {\bf\thesectionc.\thesubsectionc. {\kern1pt \bfit #1}}\par\vspace{5pt}}
\renewcommand{\subsubsection}[1] {\vspace{12pt}\addtocounter{subsubsectionc}{1}
        \noindent{\tenrm\thesectionc.\thesubsectionc.\thesubsubsectionc.
        {\kern1pt \tenit #1}}\par\vspace{5pt}}
\newcommand{\nonumsection}[1] {\vspace{12pt}\noindent{\tenbf #1}
        \par\vspace{5pt}}

\newcounter{appendixc}
\newcounter{subappendixc}[appendixc]
\newcounter{subsubappendixc}[subappendixc]
\renewcommand{\thesubappendixc}{\Alph{appendixc}.\arabic{subappendixc}}
\renewcommand{\thesubsubappendixc}
        {\Alph{appendixc}.\arabic{subappendixc}.\arabic{subsubappendixc}}

\renewcommand{\appendix}[1] {\vspace{12pt}
        \refstepcounter{appendixc}
        \setcounter{figure}{0}
        \setcounter{table}{0}
        \setcounter{lemma}{0}
        \setcounter{theorem}{0}
        \setcounter{corollary}{0}
        \setcounter{definition}{0}
        \setcounter{equation}{0}
        \renewcommand{\thefigure}{\Alph{appendixc}.\arabic{figure}}
        \renewcommand{\thetable}{\Alph{appendixc}.\arabic{table}}
        \renewcommand{\theappendixc}{\Alph{appendixc}}
        \renewcommand{\thelemma}{\Alph{appendixc}.\arabic{lemma}}
        \renewcommand{\thetheorem}{\Alph{appendixc}.\arabic{theorem}}
        \renewcommand{\thedefinition}{\Alph{appendixc}.\arabic{definition}}
        \renewcommand{\thecorollary}{\Alph{appendixc}.\arabic{corollary}}
        \renewcommand{\theequation}{\Alph{appendixc}.\arabic{equation}}
        \noindent{\tenbf Appendix \theappendixc #1}\par\vspace{5pt}}
\newcommand{\subappendix}[1] {\vspace{12pt}
        \refstepcounter{subappendixc}
        \noindent{\bf Appendix \thesubappendixc. {\kern1pt \bfit #1}}
        \par\vspace{5pt}}
\newcommand{\subsubappendix}[1] {\vspace{12pt}
        \refstepcounter{subsubappendixc}
        \noindent{\rm Appendix \thesubsubappendixc. {\kern1pt \tenit #1}}
        \par\vspace{5pt}}

\newcommand{\textlineskip}{\baselineskip=13pt}
\newcommand{\smalllineskip}{\baselineskip=10pt}

\def\eightcirc{
\begin{picture}(0,0)
\put(4.4,1.8){\circle{6.5}}
\end{picture}}
\def\eightcopyright{\eightcirc\kern2.7pt\hbox{\eightrm c}}

\def\abstracts#1#2#3{{
        \centering{\begin{minipage}{4.5in}\baselineskip=10pt\footnotesize
        \parindent=0pt #1\par 

        \parindent=15pt #2\par
        \parindent=15pt #3
        \end{minipage}}\par}}

\newcommand{\bibit}{\nineit}

\renewenvironment{thebibliography}[1]
        {\frenchspacing
         \ninerm\baselineskip=11pt
         \begin{list}{\arabic{enumi}.}
        {\usecounter{enumi}\setlength{\parsep}{0pt}
         \setlength{\leftmargin 12.7pt}{\rightmargin 0pt} 
         \setlength{\itemsep}{0pt} \settowidth
        {\labelwidth}{#1.}\sloppy}}{\end{list}}

\newcounter{itemlistc}
\newcounter{romanlistc}
\newcounter{alphlistc}
\newcounter{arabiclistc}
\newenvironment{itemlist}
        {\setcounter{itemlistc}{0}
         \begin{list}{$\bullet$}
        {\usecounter{itemlistc}
         \setlength{\parsep}{0pt}
         \setlength{\itemsep}{0pt}}}{\end{list}}

\newcommand{\fcaption}[1]{
        \refstepcounter{figure}
        \setbox\@tempboxa = \hbox{\footnotesize Fig.~\thefigure. #1}
        \ifdim \wd\@tempboxa > 5in
           {\begin{center}
        \parbox{5in}{\footnotesize\smalllineskip Fig.~\thefigure. #1}
            \end{center}}
        \else
             {\begin{center}
             {\footnotesize Fig.~\thefigure. #1}
              \end{center}}
        \fi}

\newcommand{\tcaption}[1]{
        \refstepcounter{table}
        \setbox\@tempboxa = \hbox{\footnotesize Table~\thetable. #1}
        \ifdim \wd\@tempboxa > 5in
           {\begin{center}
        \parbox{5in}{\footnotesize\smalllineskip Table~\thetable. #1}
            \end{center}}
        \else
             {\begin{center}
             {\footnotesize Table~\thetable. #1}
              \end{center}}
        \fi}

\def\@citex[#1]#2{\if@filesw\immediate\write\@auxout
        {\string\citation{#2}}\fi
\def\@citea{}\@cite{\@for\@citeb:=#2\do
        {\@citea\def\@citea{,}\@ifundefined
        {b@\@citeb}{{\bf ?}\@warning
        {Citation `\@citeb' on page \thepage \space undefined}}
        {\csname b@\@citeb\endcsname}}}{#1}}

\newif\if@cghi
\def\cite{\@cghitrue\@ifnextchar [{\@tempswatrue
        \@citex}{\@tempswafalse\@citex[]}}
\def\citelow{\@cghifalse\@ifnextchar [{\@tempswatrue
        \@citex}{\@tempswafalse\@citex[]}}
\def\@cite#1#2{{$\null^{#1}$\if@tempswa\typeout
        {IJCGA warning: optional citation argument 

        ignored: `#2'} \fi}}
\newcommand{\citeup}{\cite}

\def\pmb#1{\setbox0=\hbox{#1}
        \kern-.025em\copy0\kern-\wd0
        \kern.05em\copy0\kern-\wd0
        \kern-.025em\raise.0433em\box0}

\def\fnt#1#2{\footnotetext{\kern-.3em
        {$^{\mbox{\scriptsize #1}}$}{#2}}}

\def\fpage#1{\begingroup
\voffset=.3in
\thispagestyle{empty}\begin{table}[b]\centerline{\footnotesize #1}
        \end{table}\endgroup}

\def\runninghead#1#2{\pagestyle{myheadings}
\markboth{{\protect\footnotesize\it{\quad #1}}\hfill}
{\hfill{\protect\footnotesize\it{#2\quad}}}}
\font\tenrm=cmr10
\font\tenit=cmti10 

\font\tenbf=cmbx10
\font\bfit=cmbxti10 at 10pt
\font\ninerm=cmr9
\font\nineit=cmti9

\font\eightrm=cmr8

\def\qed{\hbox{${\vcenter{\vbox{                        
   \hrule height 0.4pt\hbox{\vrule width 0.4pt height 6pt
   \kern5pt\vrule width 0.4pt}\hrule height 0.4pt}}}$}}

\renewcommand{\thefootnote}{\fnsymbol{footnote}}        

\def\be{\begin{equation}}
\def\ee{\end{equation}}
\def\winf{W_{1+\infty}\ }
\def\u1{\widehat{U(1)}}
\def\su2{\widehat{SU(2)}_1}
\def\suem{\widehat{SU(m)}_1}

\begin{document}

\runninghead{$\winf$ Theories in the QHE} 
{$\winf$ Theories in the QHE}  

\normalsize\textlineskip
\thispagestyle{empty}
\setcounter{page}{1}

\vspace*{-80pt}
\rightline{DFF 257/10/96}
\rightline{hep-th/9610019}

\vspace*{0.88truein}

\fpage{1}
\centerline{\bf $\winf$ FIELD THEORIES}
\vspace*{0.035truein}
\centerline{\bf FOR THE EDGE EXCITATIONS}
\vspace*{0.035truein}
\centerline{\bf IN THE QUANTUM HALL EFFECT\footnote{
To appear in the proceedings of the 1996 Summer Telluride workshop
on {\it Low Dimensional Field Theory}, G. Dunne ed., World Scientific
(Singapore).}}
\vspace*{0.37truein}
\centerline{\footnotesize ANDREA CAPPELLI}
\vspace*{0.015truein}
\centerline{\footnotesize\it I.N.F.N. and Department of Physics, 
University of Florence, Largo E. Fermi 2}
\baselineskip=10pt
\centerline{\footnotesize\it I-50125 Florence, Italy}
\vspace*{10pt}
\centerline{\footnotesize CARLO A. TRUGENBERGER}
\vspace*{0.015truein}
\centerline{\footnotesize\it Department of Theoretical Physics, University 
of Geneva, 24, quai E. Ansermet}
\baselineskip=10pt
\centerline{\footnotesize\it CH-1211 Geneva 4, Switzerland}
\vspace*{10pt}
\centerline{\footnotesize GUILLERMO R. ZEMBA}
\vspace*{0.015truein}
\centerline{\footnotesize\it C.N.E.A. and Instituto Balseiro, 
Universidad Nacional de Cuyo, Centro At\'omico Bariloche}
\baselineskip=10pt
\centerline{\footnotesize\it 8400 - San Carlos de Bariloche,
R\'{\i}o Negro, Argentina}
\vspace*{0.225truein}

\vspace*{0.21truein}
\abstracts{
We briefly review these low-energy effective theories for the quantum
Hall effect, with emphasis and language familiar to field theorists. 
Two models have been proposed for describing
the most stable Hall plateaus (the Jain series):
the multi-component Abelian theories and the minimal $\winf$ models. 
They both lead to a-priori classifications of quantum Hall 
universality classes.
Some experiments already confirmed the basic predictions common to both 
effective theories, while other experiments
will soon pin down their detailed properties and differences.
Based on the study of partition functions, we show that the
Abelian theories are rational conformal field theories while the 
minimal $\winf$ models are not.}{}{}

\vspace*{1pt}\textlineskip      
\section{Introduction: the Incompressible Fluid, the Edge Excitations
and the $w_\infty$ Symmetry}
\vspace*{-0.5pt}
\noindent
In the quantum Hall effect\cite{prange}, the {\it effective field theory} 
approach has been developed to a rather high degree of sophistication. 
Its basic hypothesis are indeed fulfilled:  
the values of the Hall conductivity $\sigma_{xy}$
at the plateaus ($\sigma_{xx}=0$) display {\it universality}; moreover,
the nature of the ground state and of the low-energy excitations have been 
understood, together with their characteristic symmetry. 
Laughlin\cite{laugh} made the fundamental observation that 
the electrons form an {\it incompressible fluid}: this means that the  
density is uniform inside the sample {\it (fluid character)} and that
density waves have a gap {\it (incompressibility)}.
Actually, the latter property implies the absence of 
longitudinal conduction. 

\textheight=7.8truein
\setcounter{footnote}{0}
\renewcommand{\thefootnote}{\alph{footnote}}

Most of the properties of the incompressible fluid can be understood
at the (semi)-classical level.\citeup{laugh} \cite{ctz4}
A {\it droplet} of classical incompressible fluid is defined by the 
density function,
\be
\rho (z, \bar z, t) = \rho_0 \ \chi _{S_A(t)}\ ,\qquad
A = {N\over \rho_0}\ ,
\label{rho}\ee
where $\chi_{S_A(t)}$ is the characteristic function for a surface
$S_A(t)$ of area $A$, $N$ is the particle number, and $z=x+iy$, 
$\bar z= x-iy$ are complex coordinates on the plane.
This description is valid for energies far below the gap for density
waves, so that the density is uniform $(\rho_0)$ and the area of the droplet
is {\it constant}.
A convenient measure of the density in the quantum mechanical
problem is given by the {\it filling fraction} $\nu$,
\be
\nu= {N \over {\cal D}_A }\ , \qquad {\cal D}_A = {BA \over hc/e}\ ;
\label{filfrac}\ee
{${\cal D}_A$} is the number of one-particle states in a
Landau level, which is given by the magnetic flux through the surface 
in units of the flux quantum $\phi_o=hc/e$.
The simplest example of incompressible fluid is given by the
completely filled first Landau level ($\nu=1$), where the gap for
density waves is of the order of the cyclotron energy $\omega_c=eB/mc$
separating two Landau levels. 

\begin{figure}[t]
\vspace*{40pt}
\unitlength=0.7pt
\begin{picture}(500.00,440.00)(-10.00,0.00)
\put(0.00,500.00){\line(1,0){500.00}}
\put(0.00,80.00){\line(1,0){500.00}}
\put(250.00,500.00){\line(0,-1){100.00}}
\put(125.00,500.00){\line(0,-1){100.00}}
\put(83.00,500.00){\line(0,-1){100.00}}
\put(250.00,80.00){\line(0,1){40.00}}
\put(125.00,80.00){\line(0,1){100.00}}
\put(83.00,80.00){\line(0,1){100.00}}
\put(166.00,460.00){\makebox(0,0)[cc]{$\ \ \ \bullet {\bf 1\over 3}$}}
\put(333.00,460.00){\makebox(0,0)[cc]{$\ \ \ \bullet {\bf 2\over 3}$}}
\put(100.00,420.00){\makebox(0,0)[cc]{$\ \ \ \bullet {\bf 1\over 5}$}}
\put(200.00,420.00){\makebox(0,0)[cc]{$\ \ \ \bullet {\bf 2\over 5}$}}
\put(300.00,420.00){\makebox(0,0)[cc]{$\ \ \ \bullet {\bf 3\over 5}$}}
\put(400.00,420.00){\makebox(0,0)[cc]{$\ \ \ \bullet {\it4\over 5}$}}
\put(71.00,380.00){\makebox(0,0)[cc]{$\ \ \ \circ    {\bf 1\over 7}$}}
\put(143.00,380.00){\makebox(0,0)[cc]{$\ \ \ \bullet {\bf 2\over 7}$}}
\put(214.00,380.00){\makebox(0,0)[cc]{$\ \ \ \bullet {\bf 3\over 7}$}}
\put(286.00,380.00){\makebox(0,0)[cc]{$\ \ \ \bullet {\bf 4\over 7}$}}
\put(357.00,380.00){\makebox(0,0)[cc]{$\ \ \ \bullet {\it 5\over 7}$}}
\put(111.00,340.00){\makebox(0,0)[cc]{$\ \ \ \bullet {\bf 2\over 9}$}}
\put(222.00,340.00){\makebox(0,0)[cc]{$\ \ \ \bullet {\bf 4\over 9}$}}
\put(278.00,340.00){\makebox(0,0)[cc]{$\ \ \ \bullet {\bf 5\over 9}$}}
\put(91.00,300.00){\makebox(0,0)[cc]{$\ \ \ \circ    {\bf 2\over 11}$}}
\put(136.00,300.00){\makebox(0,0)[cc]{$\ \ \ \bullet {\bf 3\over 11}$}}
\put(182.00,300.00){\makebox(0,0)[cc]{$\ \ \ \circ   {\it 4\over 11}$}}
\put(227.00,300.00){\makebox(0,0)[cc]{$\ \ \ \bullet {\bf 5\over 11}$}}
\put(273.00,300.00){\makebox(0,0)[cc]{$\ \ \ \bullet {\bf 6\over 11}$}}
\put(318.00,300.00){\makebox(0,0)[cc]{$\ \ \ \circ   {\it 7\over 11}$}}
\put(367.00,300.00){\makebox(0,0)[cc]{$\ \ \ \circ   {\it 8\over 11}$}}
\put(115.00,260.00){\makebox(0,0)[cc]{$\ \ \ \circ   {\bf 3\over 13}$}}
\put(154.00,260.00){\makebox(0,0)[cc]{$\ \ \ \circ   {\it 4\over 13}$}}
\put(231.00,260.00){\makebox(0,0)[cc]{$\ \ \ \bullet {\bf 6\over 13}$}}
\put(269.00,260.00){\makebox(0,0)[cc]{$\ \ \ \bullet {\bf 7\over 13}$}}
\put(308.00,260.00){\makebox(0,0)[cc]{$\ \ \ \circ   {\it 8\over 13}$}}
\put(346.00,260.00){\makebox(0,0)[cc]{$\ \ \ \circ   {\it 9\over 13}$}}
\put(133.00,220.00){\makebox(0,0)[cc]{$\ \ \ \circ   {\bf 4\over 15}$}}
\put(233.00,220.00){\makebox(0,0)[cc]{$\ \ \ \circ   {\bf 7\over 15}$}}
\put(267.00,220.00){\makebox(0,0)[cc]{$\ \ \ \circ   {\bf 8\over 15}$}}
\put(235.00,180.00){\makebox(0,0)[cc]{$\ \ \ \circ   {\bf 8\over 17}$}}
\put(265.00,180.00){\makebox(0,0)[cc]{$\ \ \ \circ   {\bf 9\over 17}$}}
\put(237.00,140.00){\makebox(0,0)[cc]{$\ \ \ \circ   {\bf 9\over 19}$}}
\put(83.00,60.00){\makebox(0,0)[cc]{${\bf 1\over 6}$}}
\put(125.00,60.00){\makebox(0,0)[cc]{${\bf 1\over 4}$}}
\put(250.00,60.00){\makebox(0,0)[cc]{${\bf 1\over 2}$}}
\put(0.00,60.00){\makebox(0,0)[cc]{${\bf 0}$}}
\put(500.00,60.00){\makebox(0,0)[rc]{{\Large$\nu$}$ \qquad {\bf 1}$}}
\end{picture}
\vspace*{-10pt}
\fcaption{
Experimentally observed plateaus: their Hall conductivity 
is displayed in units of $(e^2/h)$.
The marks $\ (\bullet)\ $ denote stable (i.e. large) plateaus, which have 
been seen in several experiments; the marks $\ (\circ)\ $ 
denote less developed plateaus and plateaus found in one experiment only. 
Note that the stability decreases as the denominator of $\nu$ increases.
Coexisting fluids at the same filling fraction
have been found at $\nu=2/3,2/5,3/5,5/7$.}
\end{figure}

The Hall conductivity is simply expressed in terms of the filling fraction
\be
\sigma_{xy}={e^2 \over h }\ \nu\ ,
\label{sigmaxy}\ee
because the droplet drifts as a rigid body
under the effect of an in-plane electric field.  
The experimental values\cite{survey} show a characteristic 
{\it hierarchical pattern} (see Figure one). For the filling fractions
\be
\nu={ m\over ms \pm 1}\ ,\qquad m=1,2,\dots,\qquad s=2,4,\dots,
\label{jainfrac}\ee
Jain\cite{jain} introduced a generalization of the original Laughlin 
trial wave functions which correctly matches the microscopic dynamics.
Based on the physical picture of the {\it composite fermion}
excitation,  Jain was able to map 
strongly-interacting electrons at the filling fractions 
(\ref{jainfrac}) into weakly-interacting
composite fermions at the effective integer filling $\nu^*=m$.
By carrying over the stability of completely filled Landau levels, 
he argued that the filling fractions (\ref{jainfrac})  
form one-parameter families of approximately
equally stable Hall fluids, at fixed $s$, which accumulate at $\nu\to 1/s$.
These are called the Jain series and are represented in Figure one 
by the {\bf bold} fractions; the remaining points ({\it italic}
fractions) are less understood.
Here we shall not discuss the wave-function approach any further, but instead
show that the Jain series can be
independently derived within the effective field theory approach.

Since the droplets of incompressible fluid have constant area, 
they can only change their shape in response to external forces. 
The infinitesimal deformations are the {\it edge excitations} \cite{wen}, 
which are $(1+1)$-dimensional {\it chiral} waves.
Another type of excitations are the classical {\it vortices} in the bulk of
the droplet, which are localized holes or dips in the density.
The absence of density waves implies that
any density excess or defect is completely transmitted 
to the boundary, where it is seen as a further edge deformation.
The matter displaced at the edge defines the {\it charge} of the vortices;
instead, the edge waves are neutral excitations.

The classical configuration space of a droplet of incompressible fluid,
describing edge waves and vortices, is completely spanned by the
generators of an infinite-dimensional algebra, the $w_\infty$ algebra of  
{\it area-preserving diffeomorphisms} of the two-dimensional plane.
This is the $w_\infty\ $ {\it dynamical symmetry} of 
two-dimensional incompressible fluids.\cite{sakita} \cite{ctz1}
In order to describe this symmetry,\cite{ctz4} let us 
consider the ground state density in a rotation-invariant potential,
\be
\rho_{GS}(z, \bar z)=\rho _0 \ \Theta \left( R^2-z\bar z
\right) \ .
\label{gsd} \ee
The small deformations of the droplet at constant
area can be generated by repara\-metri\-zations of the coordinates
of the plane with unit Jacobian, the area-preserving diffeomorphisms.
A familiar example of these reparametrizations is given by the 
canonical transformations of a two-dimensional {\it phase space}.
If we identify the $(z,\bar{z})$ coordinate plane with a phase space, 
we can describe the area-preserving diffeomorphisms in terms of
canonical transformations.
Define the dimensionless Poisson brackets
$\ \{f, g\} \equiv i\ \left( \partial f
\bar \partial g - \bar \partial f \partial g \right)/\rho_0 \ $, 
where $\partial \equiv \partial / \partial z\ $ and
$\ \bar \partial \equiv \partial /\partial \bar z$.
Thus, the area-preserving diffeomorphisms are given by
$\delta z\ =\ \{{\cal L}^{(cl)}, z\}\ $ and  
$\delta {\bar z}\ =\ \{{\cal L}^{(cl)}, {\bar z}\}\ $,
in terms of the generating function
${\cal L}^{(cl)}(z,\bar z)$ of both ``coordinate'' and ``momentum''.
The basis of generators 
${\cal L}^{(cl)}_{n,m}\equiv \rho_0 ^{{n+m}/ 2} \ z^n \bar z^m \ $
satisfy the $w_{\infty }$ algebra\cite{ctz1},
\be
\left\{ {\cal L}^{(cl)}_{n,m}, {\cal L}^{(cl)}_{k,l} \right\}
=-i\ (mk-nl)\ {\cal L}^{(cl)}_{n+k-1,m+l-1} \ . 
\label{clw} \ee

The {\it small} excitations above the ground state are thus given by
the infinitesimal $w_\infty$ transformations of $\rho_{GS}$ in (\ref{rho}), 
namely by $\delta \rho _{n,m} \equiv \left\{ {\cal L}^{(cl)}_{n,m},
\rho _{GS} \right\} \ $. 
By using the Poisson brackets, one finds that 
$\delta \rho _{n,m} $ is indeed localized at the edge and that it
describes a chiral wave of momentum $k=n-m$ (the Fourier mode on the circle).
We can also consider other ``ground state'' droplets with a 
given vorticity in the bulk, and then construct the
corresponding basis of edge waves. Thus, the whole
configuration space of the excitations of a classical incompressible fluid
is spanned by infinitesimal $w_{\infty }$ transformations, as we anticipated. 

The classical edge waves and vortices have quantum analogues
in the Laughlin theory\cite{laugh}, which describes the filling 
fractions $\nu=1,1/3,1/5,\dots$.
In the simplest case of $\nu=1$, the droplet of 
electron fluid is actually a filled Fermi sea\cite{stone}:
the edge waves are particle-hole excitations across the Fermi surface 
represented by the edge of the droplet, and
the vortices correspond to localized quasi-particle and quasi-hole excitations
in the bulk of the fluid. For fractional fillings, the quantum
incompressible fluid is not easy to understand at the microscopic
level: nevertheless, Laughlin\cite{laugh} showed that the quasi-particles 
possess {\it fractional} charge and statistics.

In the quantum theory, there is also a relation\cite{ctz1} between the edge 
excitations and the generators of the quantum version of $w_{\infty }$, called
$W_{1+\infty }$.
This algebra is obtained by replacing the Poisson brackets
with quantum commutators: $i\{\ ,\ \}\ \to\ [\ ,\ ]\ $,
and by taking the thermodynamic limit of large droplets\cite{cdtz1}.
In this limit, the radius of the droplet grows as
$R\propto \ell\sqrt{N}\to\infty$, where $\ell=\sqrt{2\hbar c/eB}$ is the
magnetic length. 
Quantum edge excitations, instead, are confined
to a boundary annulus of finite size $O(\ell )$. 
Therefore, for $\nu=1$, the Fermi sea can be approximated by the
relativistic Dirac sea of a Weyl fermion.\cite{cdtz1}
The $\winf$ generators can be obtained canonically in this theory,
and satisfy the $\winf$ algebra,\cite{kac}
\be
\left[\ V^i_n, V^j_m\ \right] =(jn-im) \ V^{i+j-1}_{n+m} +
q(i,j,m,n) \ V^{i+j-3}_{n+m} +\dots +  c\ d^i(n) \ \delta ^{i,j}
\delta _{n+m,0} \ . 
\label{Win} \ee
In this equation, $i+1 \ge 1$ is the ``conformal spin'' of the generator
$V^i_n$ and $-\infty <n< +\infty $ is the edge momentum.
The first term on the right-hand-side of (\ref{Win}) reproduces the classical
$w_{\infty }$ algebra (\ref{clw}) by the correspondence
${\cal L}^{(cl)}_{i-n,i} \to V^i_n\ $ and identifies $W_{1+\infty}$ as
the algebra of ``quantum area-preserving diffeomorphisms''.
Moreover, there are some quantum corrections with polynomial
coefficients $q(i,j,n,m)$, and the $c$-number term $c\ d^i(n)$ due to the 
(relativistic) quantum anomaly.\cite{bpz} All terms in the commutation
relations (\ref{Win}) are known and uniquely determined by the closure of 
the algebra; only the {\it central charge}
$c$ is a free parameter ($c=1$ for the Weyl fermion).
The generators $V^0_n$ are Fourier modes of the fermion
density evaluated at the edge $\vert z\vert =R$, thus
$V^0_0$ measures the edge charge; instead, $V^1_0$ measures
the angular momentum of edge excitations.\cite{cdtz1}
These generators obey the Abelian current algebra $\u1$ and 
the Virasoro algebra, respectively\cite{bpz}, which 
are important sub-algebras of the $\winf$ algebra. 

The presence of a dynamical symmetry allows for a more abstract
description of the Weyl fermion theory:\cite{ctz3}
instead of writing the Hamiltonian and then proceed via canonical
quantization, we can obtain the Hilbert space by assembling the 
set of irreducible, highest-weight representations of the $W_{1+\infty}$
algebra, which is closed under the {\it fusion rules} for making composite
excitations (this is the general procedure for building
conformal field theories\cite{bpz}). Any $\winf$ representation contains 
a bottom state, the {\it highest weight} state
$\vert Q\rangle$, and an infinite tower of particle-hole excitations above it.
The bottom state represents the droplet ground-state (\ref{gsd}) $(Q=0)$, 
which might have quasi-particles in it $(Q\neq 0)$.
The charge $Q$ and the spin $J$ of the quasi-particle
are given by the eigenvalues
$V^0_0\ \vert\ Q\ \rangle\ =\ Q\ \vert\ Q\ \rangle $ and
$\ V^1_0\ \vert\ Q\ \rangle\ =\ J\ \vert\ Q\ \rangle $, while
the fractional statistics $\theta/\pi$ is twice the spin $J$.


\section{Classification of QHE Universality Classes by the $\winf$ 
Symmetry: Minimal and Non-Minimal Models}
\noindent
We have seen that the $w_\infty$ dynamical symmetry describes
the geometry of classical incompressible fluids and that,
for $\nu=c=1$, the Hilbert space for quantum
edge excitations is similarly made of $\winf$ representations. 
Actually, the effective quantum field theories of incompressible fluids 
are characterized by the $\winf$ symmetry for general filling fractions, 
because there is a unique quantization of the $w_\infty$ algebra 
in $(1+1)$-dimensional field theory.\cite{kac} 
We can therefore {\it postulate} that {\it all} universality classes 
of quantum Hall incompressible fluids are in one-to-one correspondence with
$W_{1+\infty }$ theories.\cite{ctz3}
These classes are specified by the {\it kinematical data}
of the charges $Q$ and fractional statistics $\theta/\pi$ of
the quasi-particles (the eigenvalues of $V^0_0$ and $V^1_0$). 
Moreover, the Hall conductivity (\ref{sigmaxy}) can be obtained from the 
chiral anomaly of the $(1+1)$-dimensional theory.\cite{cdtz1} \cite{ctz3}
One can also consider 
other quantities, like the characters of the $\winf$ algebra: they 
give the number of particle-hole excitations above the ground state,
which can be checked in numerical simulations of the electron fluid.\cite{wen}

This classification program can be carried through because 
all $\winf$ unitary, irreducible, highest-weight representations 
have been found.\cite{kac}
They exist for positive integer central charge 
$c=m=1,2,\dots$ and are labeled by a $m$-component highest-weight vector
$\vec{r}=\{r_1,\dots,r_m\}$, whose sum of components gives 
the charge $Q$.
The results of the representation theory can be summarized as follows:
\begin{itemlist}
\item If $c=1$, the $\winf$ representations are completely equivalent
to those of its Abelian sub-algebra $\u1$.
\item If $c=2,3,\dots$ there are two kinds of representations,
{\it generic} and {\it degenerate}, depending on the type of 
weight. The generic representations occur for 
${(r_i-r_j)} \not\in {\bf Z}\ $, $\forall\ i\neq j\ $, 
and are actually {\it equivalent} to the corresponding representations 
of the multi-component Abelian algebra $\u1^{\otimes m}$ having the same 
weight.
\item The degenerate representations have weights with some 
$(r_i-r_j) \in {\bf Z}\ $. 
These representations are {\it contained} in the corresponding
$\u1^{\otimes m}$ representations, i.e. the latter are reducible $\winf$ 
representations. The degenerate $\winf$ representations can be obtained from 
the $\u1^{\otimes m}$ ones by projecting out 
an infinity of states in the tower of excitations.
\end{itemlist}

It is rather amusing to observe that this table of mathematical
results matches the list of known effective field theories for the quantum 
Hall effect, which have been proposed by several group and were often
based on different hypotheses:
\begin{itemlist}
\item The $c=1$ Abelian theory describes the Laughlin fluids
with spectrum
\be
\nu={1\over p}\ , \quad Q={n \over p}\ ,
\quad J={n^2 \over 2p}\ , \qquad n \in {\bf Z}\ , \ \ \ 
p=1,3,5,\dots\ ,
\label{lausp}\ee
where $n$ is the number of quasi-particles.
This effective field theory is consistent with the original Laughlin 
wave-function approach and is well established.\cite{wen}
The spectrum (\ref{lausp}) and the number of edge excitations above the
ground state have been 
confirmed by numerical and real experiments (more on this later). 
An explicit action for this theory is given by the $(1+1)$-dimensional chiral 
boson (also called chiral Luttinger liquid).
This theory is equivalent, for $\nu=1$, to the Weyl fermion theory by
bosonization, and, in general, to the Abelian Chern-Simons topological
gauge theory defined on the two-dimen\-sional plane.\cite{juerg}
\item The multi-component generalization of the Abelian 
theory,\cite{juerg} \cite{zee}
carrying the $\u1^{\otimes m}$ symmetry, is the
``standard model'' for generic filling fractions. Its spectrum is
given by 
\be
Q  = \sum_{i,j=1}^m\ K^{-1}_{ij}\ n_j \ , \qquad
{ \theta\over \pi} = \sum_{i,j=1}^m\ n_i \ K^{-1}_{ij}\ n_j \ ,
\qquad \nu= \sum_{i,j=1}^m\ K^{-1}_{ij} \ .
\label{qform}\ee
Besides the central charge $c=m$, this theory is specified by a 
$(m\times m)$ symmetric, integer-valued matrix $K_{ij}$ of 
couplings (with $K_{ii}$ odd).
The $\u1^{\otimes m}$ representations in this spectrum correspond
to $c=m$ $\winf$ representations whose weights $\vec{r}_i$ are given by
the ``metric'' $K^{-1}_{ij}\sim \vec{r}_i\cdot\vec{r}_j$.
Therefore, generic forms of $K$ correspond to generic $\winf$ representations.
\item However, it turns out that the relevant Jain
filling fractions (\ref{jainfrac}) are described by the specific matrices 
$K=1+s\ C$, where $C_{ij}=1$, $\forall i,j=1,\dots, m$.
These theories are characterized\cite{juerg} \cite{zee} by the
extended symmetry $\u1\otimes\suem\supset \u1^{\otimes m}$ and,
moreover, are made of those Abelian representations which are 
{\it inequivalent} to the (degenerate) $\winf$ representations.\cite{ctz5}
\item Therefore, for the Jain filling fractions, 
{\it two} different kinds of $\winf$ theories can be built,
the {\it non-minimal} and {\it minimal} ones:\cite{ctz5} 
the Abelian theories with $\u1\otimes\suem$ symmetry are made of
reducible $\winf$ representations and are thus non-minimal;
the $\winf$ minimal models are made of irreducible degenerate representations. 
Note that all the $\winf$ minimal models were independently built 
in Ref.16 and were shown to exist for the Jain filling fractions 
{\it only}.
\end{itemlist}

In conclusion, there are two effective field theories describing
the Jain series: the Abelian theories with enhanced $\u1\otimes\suem$
symmetry and the $\winf$ minimal models.
Both theories display the same spectrum of charge and fractional 
spin of quasi-particles, but have different multiplicities of excitations.
The $\winf$ minimal models are equivalent to the 
$\u1\otimes {\cal W}_m$ conformal theories\cite{kac}, 
where ${\cal W}_m$ is the Zamolodchikov-Fateev-Lykyanov algebra.\cite{fz} 
Since the  ${\cal W}_m$ representations are isomorphic to those of the
$SU(m)$ Lie algebra, the neutral excitations of the minimal models have
associated an $SU(m)$  {\it ``isospin''} quantum number, i.e. they are 
{\it quark-like} and their statistics is {\it non-Abelian}.\cite{ctz5}
This means that the four-point functions have several intermediate
channels, which can be observed in the scattering of two quasi-particles. 
Moreover, the number of excitations above the ground state
is smaller in the $\winf$ minimal models than in the Abelian
ones, due to the corresponding inclusion of representations.
A more detailed analysis\cite{cz} shows that the minimal models have $SU(m)$
quantum numbers but cannot realize the full $SU(m)$ symmetry
(for ex., in the $SU(2)$ case, $J_+$ is present, while 
$J_-,J_0$ are missing).

Present theoretical a-priori arguments, like consistency
conditions or symmetries cannot choose between one of the two theories.
In the literature, the non-minimal $\u1\otimes\suem$ theory was first 
introduced and is more widely accepted. 
It has been argued that the $\nu=m$ plateaus possess $m$ independent edges
($m$ Weyl fermions), which naturally realize the 
$\u1\otimes\suem$ symmetry: this might extend to the
Jain plateaus by the composite-fermion correspondence.
However, the $m$ Landau levels are not equivalent at the
microscopic level, as well as in the composite-fermion theory, 
even for $\omega_c=0$.
Unfortunately, a direct relation between the Jain theory and the edge 
approach has not been found yet.
Finally, the numerical experiments counting the number of excited states 
could distinguish between the two theories, but they are not accurate enough
at present. 

\section{Experiments}
\noindent
Let us now review the experiments which have already confirmed the 
simpler $c=1$ Abelian theory of edge excitations
and discuss further experiments which could identify one of the
two proposed $c>1$ theories:
\begin{itemlist}
\item The ``time domain'' experiment\cite{tdom} has observed the
propagation of waves along the edge of a disk sample;
a short pulse was injected at one point on the edge and
detected at another edge point by a fast oscilloscope.
This confirmed that the $\nu=1/3$ Hall fluid has a single 
chiral excitation as well as the $\nu=2/3$ one (which has
a second neutral excitation according to both effective theories)
\item The characteristic low-energy spectrum of edge excitations 
$\epsilon_k\sim (k\log k)/R$ was found 
by an experiment of {\it radio-frequency} resonance.\cite{wasser}
This was not considered a proof of the edge states 
because the spatial propagation could not be resolved.
\item An interesting recent experiment is the
resonant tunneling through a point contact,
proposed in Ref.21 and first done in Ref.22.
Usually, edge excitations are chiral and do not self-interact;
however, one can consider a rectangular geometry and pinch the electron 
fluid at the mid point by applying a localized potential barrier, 
such that the two opposite edges interact.
One observes resonance peaks in the conductivity 
when a quasi-particle or an electron tunnels through the point contact.
At $\nu=1/3$, the Abelian theory predicts the anomalous temperature dependence
$T^{2/3}$ (as compared to the free Fermi liquid) of the half-width of the 
resonance peaks and a characteristic peak shape.\cite{tunn}
The first observation of these effects has not been confirmed by 
other experiments and the current belief
is that they should be seen at much lower temperatures.\cite{geller}
Nevertheless, another tunneling experiment has 
measured the fractional charge of the quasi-particles.\cite{goldman}
\end{itemlist}

Generalizations of the tunneling experiments to other
filling fractions $\ \ \nu=2/3,2/5,\dots$ will test the $c>1$
effective theories; however, their characteristic neutral
excitations have indirect effects in conduction experiments.
Moreover, an interference experiment involving the four-point
function must be devised to distinguish between the Abelian and non-Abelian 
statistics which characterize the two proposed theories.
Perhaps, this could be achieved in the tunneling experiment
with two point contacts recently proposed in Ref.25. 
In conclusion, we hope to have convinced the reader that
the effective theories of the quantum Hall effect are now being tested by
very interesting experiments.


\section{The Abelian Theories are Rational Conformal Field Theories}
\noindent
The quantum numbers of edge excitations always take rational
values; this suggests that the corresponding conformal field theories 
should belong to the special, well understood class of
{\it rational } conformal field theories (RCFT)\cite{mose}.
By definition, these contain a {\it finite} number of
representations of an extended (chiral) symmetry algebra, which includes 
Virasoro as a sub-algebra: clearly, each of the representations 
describes a sector of the Hilbert space which contains infinite states. 
These representations are encoded in the partition function\cite{bpz} 
of the Euclidean theory defined on the  
space-time torus $S^1\times S^1$,
\be
Z(\tau)=\sum_{\lambda,\bar\lambda=1}^N\ {\cal N}_{\lambda\bar\lambda}
\ \chi(\tau)_\lambda\ \overline{\chi(\tau)}_{\bar\lambda}\ ,
\label{ratz}\ee
where $\tau$ is the ratio of the two periods of the torus,
$\chi_\lambda$ are the characters of the extended algebra representations and
${\cal N}_{\lambda\bar\lambda}$ are the (unknown) multiplicities 
of the representations. 
The partition function gives a precise inventory of all the states in the
Hilbert space and serves as a {\it definition} of the RCFT. 
In the literature,\cite{bpz} it was shown that $Z$ is invariant 
under the modular transformations $\tau\to\tau +1$ and $\tau\to -1/\tau$,
which span the discrete group $\Gamma=SL(2,{\bf Z})/{\bf Z}_2$.
Furthermore, Verlinde\cite{verl} has shown the relation between 
modular invariance and the fusion rules of the extended symmetry
algebra.
Moreover, Witten\cite{jones} has explained that any RCFT is 
associated to a Chern-Simons theory, and that the torus partition function
in the former theory corresponds to a path-integral amplitude for
the latter theory on the manifold
$S^1\times S^1\times {\bf R}$, where ${\bf R}$ is the time axis.

In the quantum Hall effect, a complete description of the
edge excitations similarly requires the construction of their 
partition function.\cite{cz}
Consider a spatial annulus with Euclidean compact time:
this space-time manifold is topologically ${\cal M}=S^1\times S^1\times I$,
because the radial coordinate runs over the finite interval 
$I$ and the angular coordinate and Euclidean time are both compact.
We can define the partition function of the conformal field theory
on the edge(s) $\partial{\cal M}=S^1\times S^1$, which is a space-time 
torus.
This annulus partition function takes the standard form (\ref{ratz}),
where the characters of the chiral and antichiral algebras
pertain to the inner and outer edges, respectively.

The construction of the partition functions for the simpler
$c=1$ Abelian theories shows that these are indeed rational conformal
field theories; for $\nu=1/p$, the extended symmetry algebra is the 
extension of $\u1$ by a current of Virasoro dimension $h=p/2$, and the
Verlinde fusion rules are the Abelian group ${\bf Z}_p$ (addition modulo $p$). 
The number $p$ of extended algebra representations is equal to the Wen 
{\it topological order},\cite{topord} which is the degeneracy 
of the quantum Hall ground state on the (ideal) compact surface of the torus
- another universal property of Hall fluids.
This degeneracy is usually accounted for by the effective Chern-Simons 
theory, but it can be equivalently obtained from
the annulus partition function, thanks to the 
Verlinde and Witten relations.\cite{cz}

The partition functions for the two $c>1$ effective 
theories show new features:\cite{cz} 
\begin{itemlist} 
\item For the Abelian $\u1\otimes\suem$ theories, there are several 
solutions to the modular invariance conditions, leading to 
chiral-antichiral {\it diagonal} and {\it non-diagonal} partition functions.
\item The partition functions corresponding to the minimal $\winf$
theories cannot be modular invariant. 
\end{itemlist}

Let us first discuss the Abelian theories.
A diagonal partition function (${\cal N}_{\lambda\bar\lambda}=
\delta_{\lambda\bar\lambda}$ in Eq.(\ref{ratz})) is found for each of the
{\it chiral} theories previously discussed.
It contains the excitations in Eq.(\ref{qform}) for each edge of the annulus,
and, moreover, paired two-edge excitations.
On the other hand, the non-diagonal partition functions
define new RCFTs, which describe further observed plateaus
beyond the Jain series, as follows.
They exist for $m$ containing a square factor, $m=a^2m^\prime$, and
possess the extended symmetry algebra
$\u1\otimes\widehat{SU\left(a^2 m^\prime \right)}_1$ and a
smaller set of neutral quasi-particles.
Their filling fractions span again the Jain series, with
$m\to m^\prime$, and further series of fractions which 
include $\nu=2/3,4/5,6/7,8/9,\dots$, $4/11, 4/13$ (only) for small denominator.
The experimental points beyond the Jain series (in italics in Figure one)
match rather well these new values and 
their (less stable) ``charge-conjugated'' partners $\nu\to (1-\nu)$. 
This has to be compared with the higher orders of the Jain hierarchy
of wave functions\cite{jain} and similar constructions, which generically
predict too many unobserved filling fractions.

Next, we discuss the $\winf$ minimal models. Since they do not have a
modular invariant partition function, they are 
not rational conformal field theories and do not correspond to  
Chern-Simons theories. Therefore, they are less understood in the literature.
Clearly, they are fully consistent conformal field theories and their 
relevance is mainly an experimental question, as discussed before.
Presumably, the projection from the Abelian to the minimal theories 
(the Hamiltonian reduction\cite{hamred}) is incompatible  with modular 
invariance. Their partition functions should be found 
by a novel approach, in which modular invariance would be appropriately 
generalized, possibly along the lines of Ref.31.

In conclusions, we would like to mention the model building associated
with other quantum Hall states with spinful or layered electrons\cite{mr}
(called {\it non-Abelian} Hall fluids for historical reasons).
The simple semiclassical droplet model of section one does not
apply to these more involved fluids; thus, their effective
edge theories might fall outside the $\winf$ classification.
We remark that the modular invariant partition 
function and the associated RCFT calculus can be useful for understanding
these edge theories, as initiated in Ref.33.

\nonumsection{Acknowledgements}
\noindent
A.C. is partially supported by the European Community 
Programme ``Human Capital and Mobility''.
C.A.T. is supported by a Profil 2 fellowship of the Swiss National Science 
Foundation. G.R.Z. is supported in part
by a grant of the Antorchas Foundation (Argentina).


\def\NP{{\bibit Nucl. Phys.\ }}
\def\PRL{{\bibit Phys. Rev. Lett.\ }}
\def\PL{{\bibit Phys. Lett.\ }}
\def\PR{{\bibit Phys. Rev.\ }}
\def\CMP{{\bibit Commun. Math. Phys.\ }}
\def\IJMP{{\bibit Int. J. Mod. Phys.\ }}
\def\MPL{{\bibit Mod. Phys. Lett.\ }}
\def\RMP{{\bibit Rev. Mod. Phys.\ }}
\def\AP{{\bibit Ann. Phys. (NY)\ }}

\nonumsection{References}


\begin{thebibliography}{000}
\bibitem{prange} For a review see: R. A. Prange, S. M. Girvin, 
		{\bibit The Quantum Hall Effect} (Springer, New York, 1990).
\bibitem{laugh} for a review see: R. B. Laughlin, {\bibit Elementary Theory: 
		the Incompressible Quantum Fluid}, in Ref.\cite{prange}
\bibitem{ctz4}  A. Cappelli, C. A. Trugenberger and G. R. Zemba,
                \AP {\bibit 246} (1996) 86.
\bibitem{survey}D. C. Tsui, {\bibit Physica} {\bibit B 164} (1990) 59;
                H. L. Stormer, {\bibit Physica} {\bibit B 177} (1992) 401;
                T. Sajoto, Y. W. Suen, L. W. Engel, M.B. Santos and M. Shyegan,
                \PR {\bibit B 41} (1990) 8449.
\bibitem{jain}  For a review see: J. K. Jain, {\bibit Adv. in Phys.} 
		{\bibit 41} (1992) 105, {\bibit Science} {\bibit 266} 
		(1994) 1199.
\bibitem{wen}   for a review, see: X. G. Wen, \IJMP {\bibit B 6} (1992) 1711.
\bibitem{sakita}S. Iso, D. Karabali and B. Sakita,
                \NP {\bibit B 388} (1992) 700, \PL {\bibit B 296} (1992) 143.
\bibitem{ctz1}  A. Cappelli, C. A. Trugenberger and G. R. Zemba,
                \NP {\bibit B 396} (1993) 465, \PL {\bibit B306} (1993) 100;
                for a review, see:
                A.Cappelli, G.V.Dunne, C.A.Trugenberger and G.R.Zemba,
                \NP {\bibit B (Proc. Suppl.) 33C} (1993) 21.
\bibitem{stone} M. Stone, \AP {\bibit 207} (1991) 38.
\bibitem{cdtz1} A. Cappelli, G. V. Dunne, C. A. Trugenberger and G. R.
                Zemba, \NP {\bibit B 398 } (1993) 531.
\bibitem{kac}  V. Kac and A. Radul, {\bibit Comm. Math. Phys.} {\bibit 157}
                (1993) 429; H. Awata, M. Fukuma, Y. Matsuo and
                S. Odake, {\bibit Prog. Theor. Phys. (Suppl.)} {\bibit 118}
                (1995) 343; E. Frenkel, V. Kac, A. Radul and W. Wang,
                \CMP {\bibit 170} (1995) 337; V. Kac and A. Radul,
                hep-th/9512150.
\bibitem{bpz}   for a review see: P. Ginsparg, in {\bibit Fields, Strings 
		and Critical Phenomena},
                Les Houches School 1988, E. Brezin and J. Zinn-Justin eds.
                (North-Holland, Amsterdam, 1990).
\bibitem{ctz3}  A. Cappelli, C. A. Trugenberger and G. R. Zemba,
                \PRL {\bibit 72} (1994) 1902.
\bibitem{juerg} J. Fr\"ohlich and A. Zee, \NP {\it B 364 } (1991) 517. 
\bibitem{zee}   X.-G. Wen and A. Zee, \PR {\bibit B 46} (1993) 2290;
                J. Fr\"ohlich, T. Kerler, U. M. Studer 
                and E. Thiran, \NP {\bibit B 453} (1995) 670.
\bibitem{ctz5}  A. Cappelli, C. A. Trugenberger and G. R. Zemba,
                \NP {\bibit 448} (1995) 470; for a review, see: 
                \NP {\bibit (Proc. Suppl.)} {\bibit B 45A} (1996) 112.
\bibitem{fz}    V. A. Fateev and A. B. Zamolodchikov, \NP {\bibit B 280} 
		(1987) 644; V. A. Fateev and S. L. Lykyanov, \IJMP {\it A 3} 
		(1988) 507.
\bibitem{cz}    A. Cappelli and G. R. Zemba, {\bibit Modular Invariant
		Partition Functions in the Quantum Hall Effect}, 
		hep-th/9506127; see also T. Gannon, hep-th/9608063.
\bibitem{tdom}  R. C. Ashoori, H. L. Stormer, L. N. Pfeiffer, K. W. Baldwin
                and K. West, \PR {\bibit B 45} (1992) 3894;
\bibitem{wasser}M. Wassermeier, J. Oshinowo, J. P. Kotthaus, A. H.
		MacDonald, C. T. Foxon and J. J. Harris, \PR {\bibit B 41}
		(1990) 10287.
\bibitem{tunn}  K. Moon, H. Yi, C. L. Kane, S. M. Girvin and M. P. A. Fisher,
                \PRL {\bibit 71} (1993) 4381; P. Fendley, A. W. W. Ludwig and
                H. Saleur, \PRL {\bibit 74} (1995) 3005, \PR {\bibit B 52} 
                (1995) 8934, cond-mat/9601117.
\bibitem{milli} F. P. Milliken, C. P. Umbach and R. A. Webb, 
                {\bibit Solid State Commun.} {\bibit 97} (1996) 309.
\bibitem{geller}M. R. Geller, D. Loss and G. Kirczenow, cond-mat/9606070.
\bibitem{goldman}I. J. Maasilta and V. J. Goldman, {\bibit Lineshape of 
		Resonant Tunneling between Fractiona Quantum Hall Edges},
		Stony-Brook preprint 1996.
\bibitem{tunn2} C. de C. Chamon, D. E. Freed, S. A. Kivelson, S. L. Sondhi 
		and X. G. Wen, cond-mat/9607195.
\bibitem{mose}  for a review, see: G. Moore and N. Seiberg, {\bibit Lectures on
                RCFT}, proceedings of the 1989 Banff Summer school, 
                H. C. Lee ed. (Plenum Press, New York, 1990).
\bibitem{verl}  E. Verlinde, \NP {\bibit B300} (1988) 360.
\bibitem{jones} E. Witten, \CMP {\bibit 121} (1989) 351.
\bibitem{topord}X.-G. Wen, \PR {\bibit 40} (1989) 7387, \IJMP {\bibit B 4} 
		(1990) 239, \IJMP {\bibit B 5} (1991) 1641; for a review, see:
                {\bibit Quantum Hall Effect}, M. Stone ed. (World Scientific,
                Singapore, 1992).
\bibitem{hamred}M. Bershadsky and H. Ooguri, \CMP {\bibit 126} (1989) 49; 
		V. A. Fateev and S. L. Lykyanov, \IJMP {\bibit A 7} (1992) 
		853, 1325. 
\bibitem{dijk}  R. Dijkgraaf, hep-th/9609022.
\bibitem{mr}    G. Moore and N. Read, \NP {\bibit B 360} (1991) 362;
		X. G. Wen, \PRL {\bibit 70} (1993) 355;
		X. G. Wen and Y. S. Wu, \NP {\bibit B 419} (1994) 455;
                X. G. Wen, Y. S. Wu, and Y. Hatsungai, \NP {\bibit B 422} 
                (1994) 476.
\bibitem{read}  E. Keski-Vakkuri and X.-G. Wen, \IJMP {\bibit B 7} (1993) 4227;
		M. Milovanovi\'c and N. Read, cond-mat/9602113.
\end{thebibliography}
\end{document}